\documentclass[11pt]{article}
\usepackage{amsfonts}
\usepackage{amsmath,amssymb}
\usepackage{epsfig}
\def\dref#1{(\ref{#1})}

\newlength{\gH}
\newlength{\gV}
\begin{document}

\begin{center}
{{\Large \bf $L_2$ norm performance index of synchronization and
optimal control synthesis of complex networks}}
 \footnote{{\footnotesize This work is supported by both the National Science
 Foundation of China under grants 60674093, 60334030 and

\quad the City University of Hong Kong under the Research
Enhancement Scheme and SRG grant 7002134. }}
\end{center}
\vskip 0.3cm
\begin{center}
  Chao Liu $^{\textrm{a}}$ \footnote{
{\footnotesize Corresponding author: chaoliu@pku.edu.cn}},
    \quad Zhisheng Duan $^{\textrm{a}}$,
    \quad Guanrong Chen $^{\textrm{b}}$,
    \quad Lin Huang $^{\textrm{a}}$
\end{center}
\begin{center}
{\small \it  $^{\textrm{a}}$ State Key Laboratory for Turbulence and
Complex Systems, Department of Mechanics and \\Aerospace
Engineering, College of Engineering, Peking University, Beijing
100871, P. R. China
\\ $^{\textrm{b}}$ Department of Electronic Engineering, City University of Hong Kong, Hong Kong, SAR, P. R. China}
%\\{\it Email: chaoliu@pku.edu.cn}
\end{center}

\vspace*{1\baselineskip}
\begin{center}
{\begin{minipage}{127mm} {\small {\bf Abstract.} In this paper, the
synchronizability problem of dynamical networks is addressed, where
better synchronizability means that the network synchronizes faster
with lower-overshoot. The $L_2$ norm of the error vector $e$ is
taken as a performance index to measure this kind of
synchronizability. For the equilibrium synchronization case, it is
shown that there is a close relationship between the $L_2$ norm of
the error vector $e$ and the $H_2$ norm of the transfer function $G$
of the linearized network about the equilibrium point. Consequently,
the effect of the network coupling topology on the $H_2$ norm of the
transfer function $G$ is analyzed. Finally, an optimal controller is
designed, according to the so-called \textit{LQR} problem in modern
control theory, which can drive the whole network to its equilibrium
point and meanwhile minimize the $L_2$ norm of the output of the
linearized network. }

\vspace*{0.5\baselineskip} {{\bf Keywords.} network,
 synchronizability, $L_2$ norm, $H_2$ norm, \textit{LQR} problem, optimal control.}

\end{minipage}
}
\end{center}

\abovedisplayskip=0.13cm \abovedisplayshortskip=0.06cm
\belowdisplayskip=0.13cm \belowdisplayshortskip=0.08cm

\def\dis{\displaystyle}

%\vspace*{3\baselineskip}

\section{Introduction}

\quad The topic of synchronizability of dynamical networks has
attracted increasing interest recently (see
\cite{bern05,hong04,nish03,wu03,wu05,wu06,zhou06} and references
therein), which is mostly referred to as how easy it is for the
network to synchronize. Technically, this is a problem whether there
is a wide range of coupling strength in which the synchronization is
stable, and the wider the range, the better the synchronizability of
the network. Accordingly, for some cases when the synchronized
region is unbounded, a better synchronizability means that the
synchronization can be achieved with a smaller coupling strength
\cite{wu03}. Generally, the synchronizability of a network depends
on the underlying coupling topology of the network. Studies show
that the ability for a network to synchronize is related to the
spectral properties of the outer coupling matrix of the network
\cite{wu03,wu05} and thus influenced by the structural properties of
the network, such as average distance \cite{zhou06}, degree
homogeneity \cite{nish03}, clustering coefficient \cite{wu06},
degree correlation \cite{bern05}, betweenness centrality
\cite{hong04}, etc.

In this paper, the synchronizability problem of dynamical networks
is considered from a different point of view: suppose that the
coupling strength of a network belongs to the range in which the
synchronization is stable; then, how fast the synchronization will
be achieved? This problem is important for the reason that in
practical engineering implementation (such as communications via
chaotic synchronization), synchronization is expected to be achieved
not only easily but also swiftly. In \cite{nish06}, it is
qualitatively pointed out that networks with diagonalizable outer
coupling matrices may synchronize faster than the ones with
non-diagonalizable outer coupling matrices. Clearly, a measure is
needed for a quantitative description of the swiftness of network
synchronization. To meet this objective, the $L_2$ norm of the error
vector $e$, denoted as $\|e\|_2$, is taken in this paper as a
performance index of this kind of network synchronizability. As will
be seen later, the quantity $\|e\|_2$ presents a suitable measure of
both swiftness and overshoot (referring to the largest difference
among various node dynamics before the synchronization is achieved):
the smaller the quantity $\|e\|_2$, the faster with smaller
overshoot the network synchronization.

Furthermore, as shown by the numerical examples given below, the
quantity $\|e\|_2$ is influenced by the coupling topology of the
network. Thus, the investigation on the relationship between
$\|e\|_2$ and the network structure is of significance. In this
paper, for the case that the synchronous state is an equilibrium
point, it is pointed out that $\|e\|_2$ is upper-bounded by the
product of the vector 2-norm of the initial error vector $e_0$ and
the $H_2$ norm of the transfer function $G(s)$, denoted as
$\|G(s)\|_2$ or simply $\|G\|_2$, of the linearized network about
the equilibrium point. Thus, the smaller the $\|G\|_2$, the smaller
the $\|e\|_2$ as well. As pointed out in \cite{duan07}, the
relationship between $\|G(s)\|_2$ and the network structure is quite
complicated. Under some assumptions, it is proved in this paper (see
Theorem 1 and Example 4) that $\|G\|_2$ will not increase as the
real eigenvalues of the symmetrical outer coupling matrix increase.

For a linear time-invariant system, the linear quadratic regulator
problem, or simply the \textit{LQR} problem, is a classical problem
in modern control theory. The objective of the \textit{LQR} problem
is to find an optimal control law $u(t)$ such that the state $x(t)$
is driven into a (small) neighborhood of the origin while minimizing
a quadratic performance ($L_2$ performance) index on $u$ and $x$. In
fact, the \textit{LQR} problem is posed traditionally as the
minimization problem of the $L_2$ norm of the regulator output of
the system. In this paper, based on the techniques of the
\textit{LQR} problem, an optimal controller design is developed so
as to drive the network dynamics onto some homogenous stationary
states while minimizing the $L_2$ norm of the output of the
linearized network.

The rest of the paper is organized as follows. In Section 2, some
preliminary definitions and lemmas necessary for successive
development are presented. In Section 3, some numerical examples are
provided to illustrate that the quantity $\|e\|_2$ presents a
suitable measure of both swiftness and overshoot of the network
synchronization. For the equilibrium synchronization case, the
relationship between $\|e\|_2$ and the network structure is
investigated in Section 4. Based on the results of the \textit{LQR}
problem, an \textit{LQR} optimal controller is proposed in Section
5. The paper is concluded by the last Section.

\section{Preliminaries}

\quad $L_2[a,b]$ is an infinite-dimensional Hilbert space, which
consists of all square-integrable and Lebesgue measurable functions
defined on an interval $[a,b]$ with the scalar inner product
$$
\langle f,g \rangle=\int_{a}^{b}f(t)^*g(t)dt,
$$
while if the functions are vector or matrix-valued, the inner
product is defined as
$$
\langle f,g \rangle=\int_{a}^{b}\textrm{trace}[f(t)^*g(t)]dt,
$$
where $\ast$ denotes complex conjugate transpose, and the induced
norm is defined as
\begin{equation}\label{L2norm}
\|f\|_2=\sqrt{\langle f,f \rangle},
\end{equation}
for $f,\;g\in L_2[a,b]$.
%In this paper, we will focus on the Hilbert
%space $L_2[0,+\infty)$.

Consider a continuous-time linear system,
\begin{equation}\label{pre1}
\dot{x}=Ax+Bu,\quad y=Cx,
\end{equation}
%where $x$ is the state of the system, $u$ and $y$ are the input and
%output of the system, respectively,
where $x$, $u$ and $y$ are the state, input and output of the
system, respectively, and $A\in \mathbb{R}^{n\times n}$, $B\in
\mathbb{R}^{n\times m}$ and $C\in \mathbb{R}^{l\times n}$ are given
constant matrices. The transfer function from $u$ to $y$ is
$G(s)=C(sI-A)^{-1}B$. If $A$ is stable, then the $H_2$ norm of
system \dref{pre1} is represented by the $H_2$ norm of the transfer
function $G(s)$, which is defined by
$$
\|G(s)\|_2
=\sqrt{\frac{1}{2\pi}\int_{-\infty}^{+\infty}\textrm{trace}\{G^*(j\omega)G(j\omega)\}d\omega}.
$$
It can be proved that $\|G(s)\|_2^2$ equals the overall output
energy of the system response to the impulse input. For computing
$\|G(s)\|$, the following formula is convenient.
%%\cite{zhou96}.

\textbf{Lemma 1} \cite{zhou96}: If $A$ is stable, then
$\|G(s)\|_{2}^2=\textrm{trace}(B^TYB)$, where matrix $Y$ is the
solution to the following Lyapunov equation:
\begin{equation}\label{pre2}
YA+A^TY+C^TC=0.
\end{equation}
Equivalently,
\begin{equation}\label{pre3}
\|G(s)\|_{2}^2=\inf\limits_{P>0}\left\{\textrm{trace}(B^TPB):\;PA+A^TP+C^TC\leq
0\right\}.
\end{equation}

%When $A$ is unstable, the $H_2$ norm can be computed \cite{zhou96}.
The so-called linear quadratic regulator (\textit{LQR}) problem is
an optimal control problem with a quadratic performance ($H_2$ norm)
criterion. For the linear time-invariant system
$$
\dot{x}=Ax+Bu,\;x(t_0)=x_0,
$$
where $x_0$ is arbitrarily given, the regulator problem refers to
finding a control function $u(t)$ defined on $[t_0,T]$, which can be
a function of the state $x(t)$, such that the state $x(t)$ is driven
into a (small) neighborhood of origin at time $T$, $T<\infty$. Since
every physical system has energy limitation, and large control
action (even if it is realizable) can easily drive the system out of
its valid region, certain limitations have to be imposed on the
control in practical engineering implementation. For these reasons,
the regulator problem is usually posed as an optimal control problem
with a certain combined performance index on $u$ and $x$. Focusing
on the infinite time regulator problem (i.e., $T\rightarrow\infty$)
and without loss of generality assuming $t_0=0$, the \textit{LQR}
problem is formulated as follows: Find a control $u(t)$ defined on
$[0,\;\infty)$ such that the state $x(t)$ is driven to the origin at
$t\rightarrow\infty$ and the following performance index is
minimized:
\begin{equation}\label{pre4}
\min\limits_{u}\int_{0}^{\infty}\left[\begin{array}{c}x(t) \\
u(t)\end{array}\right]^*\left[\begin{array}{cc}Q & S\\
S^* & R\end{array}\right]\left[\begin{array}{c}x(t) \\
u(t)\end{array}\right]dt
\end{equation}
for some $Q=Q^*,\;S$ and $R=R^*>0$. Here, $R>0$ emphasizes that the
control energy has to be finite, i.e., $u(t)\in L_2[0,\;\infty)$.
Moreover, it is assumed that
\begin{equation}\label{pre5}
\left[\begin{array}{cc}Q & S\\
S^* & R\end{array}\right]\geq 0,\;R>0.
\end{equation}
Then, \dref{pre5} can be factored as
$$
\left[\begin{array}{cc}Q & S\\
S^* & R\end{array}\right]=\left[\begin{array}{c}C^* \\
D^*\end{array}\right]\left[\begin{array}{cc}C & D\end{array}\right]
$$
and \dref{pre4} can be written as
$$
\min\limits_{u(t)\in L_2[0,\;\infty)}\|Cx+Du\|_2^2.
$$
Traditionally, the \textit{LQR} problem is posed as the following
minimization problem:

\begin{equation}\label{pre6}
\begin{array}{l}
\min\limits_{u(t)\in L_2[0,\;\infty)}\|y\|_2^2
\\
\textrm{subject to:}
\left\{\begin{array}{l}\dot{x}=Ax+Bu,\;x(t_0)=x_0\\
y=Cx+Du.\end{array}\right.
\end{array}
\end{equation}

For the above \textit{LQR} problem, the following lemma is useful.
%%\cite{zhou96}.

\textbf{Lemma 2 } \cite{zhou96}: Suppose that in \dref{pre6}:

(A1) $(A,\;B)$ is stabilizable;

(A2) $D$ has full column rank with $[D\; D_\bot]$ being unitary;

(A3) $(C,\;A)$ is detectable;

(A4) $\left[\begin{array}{cc}A-j\omega I & B\\
C & D\end{array}\right]$ has full column rank for all $\omega$.\\
Then, there exists a unique optimal control $u=Fx$ for the
\textit{LQR} problem \dref{pre6}, where
\begin{equation}\label{pre8}
F=-(B^*X+DC)
\end{equation}
and $X$ is the stabilizing solution to the following Riccati
equation:
\begin{equation}\label{pre9}
(A-BD^*C)^*X+X(A-BD^*C)-XBB^*X+C^*D_\bot D_\bot^*C=0.
\end{equation}
Moreover, the minimized $L_2$ norm of the output $y(t)$ is given by
\begin{equation}\label{pre10}
\min\limits_{u(t)\in L_2[0,\;\infty)}\|y\|_2^2=\|G_cx_0\|_2^2,
\end{equation}
where $G_c$ is the transfer function of the system
$$
\left\{\begin{array}{l}\dot{x}=(A+BF)x+Ix_0\delta(t),\;x(0_-)=0\\
y=(C+DF)x,\end{array}\right.
$$
with $I$ being the identity matrix and $\delta(t)$ the impulse
function.

\section{$L_2$ norm performance index of network synchronization}

\quad Consider a network of $N$ identical dynamical nodes, described
by
\begin{equation}\label{net1}
\dot{x}_i=f(x_i)-\sigma\sum_{j=1}^Nm_{ij}\Gamma
x_j,\;i=1,2,\cdots,N,
\end{equation}
where $f(x)$ governs the dynamics of each individual node, $\sigma$
is the coupling strength, $\Gamma$ is the inner linking matrix, and
$M=\left[m_{ij}\right]$ is the outer coupling matrix.

Let
\begin{equation}\label{net2}
e_i(t)= x_i(t)-x_1(t),\; i=1,\cdots,N,
\end{equation}
and
\begin{equation}\label{net3}
e(t)=[e_1^T \;e_2^T \;\cdots \;e_N^T]^T
\end{equation}
denote the error vector of network \dref{net1}. Then, network
\dref{net1} is said to achieve (asymptotical) synchronization if
$$
e(t)\rightarrow 0,\;\textrm{as}\; t\rightarrow \infty.
$$

Let $\|e(t)\|_2([0,T])$, or simply $\|e(t)\|_2$ when no confusion
may be caused, denote the $L_2$ norm of $e(t)$ on a given interval
$[0,\;T]$. Then, according to \dref{L2norm},
\begin{equation}\label{net4}
%L_2(e)=
\|e(t)\|_2=\sqrt{\langle e,e
\rangle}=\sqrt{\int_{0}^{T}\textrm{trace}[e(t)^*e(t)]dt}.
\end{equation}

In fact, $\|e(t)\|_2$ represents the energy of the error signal $e$
on the interval $[0,T]$, which is in proportion to the area between
the error function $e$ and the time axis. Hence, $\|e(t)\|_2$ can be
used as a quantitative measure of the swiftness and overshoot of the
network synchronization. In what follows, two examples are first
given for illustration.

\textbf{Example 1:} Suppose that each single node in network
\dref{net1} is a Chua's oscillator. In the dimensionless form,
Chua's oscillator is described by
\begin{equation}\label{chua1}
\left\{
\begin{array}{l}\dot{x}_1=\alpha (-x_1+x_2-f(x_1)),\\
\dot{x}_2=x_1-x_2+x_3,\\
\dot{x}_3=-\beta x_2-\gamma x_3,\\
\end{array}\right.
\end{equation}
where $f(\cdot)$ is a piecewise linear function:
\begin{equation}\label{chua2}
f(x_1)=m_1x_1+\frac{1}{2}(m_2-m_1)(|x_1+1|-|x_1-1|).
\end{equation}

Take parameters $\alpha=9,\;\beta=14,\;\gamma=0.01,\;m_1=-0.714$,
and $m_2=-1.14$, so that Chua's oscillator \dref{chua1} generates a
double-scroll chaotic attractor.
%as shown in Fig. 1.
Set $\Gamma=\textrm{diag}(1,1,1)$ in \dref{net1}. Fig. 1 shows the
different synchronization performances of network \dref{net1} with
the same coupling strength $\sigma$ ($\sigma=6$) but different
coupling configurations. The corresponding values of $\|e(t)\|_2$
are computed numerically as given in Table 1.

\begin{center}
\vskip -0.5cm
 \unitlength=1cm
 \qquad \hbox{\hspace*{0.1cm} \epsfxsize5cm \epsfysize5cm
\epsffile{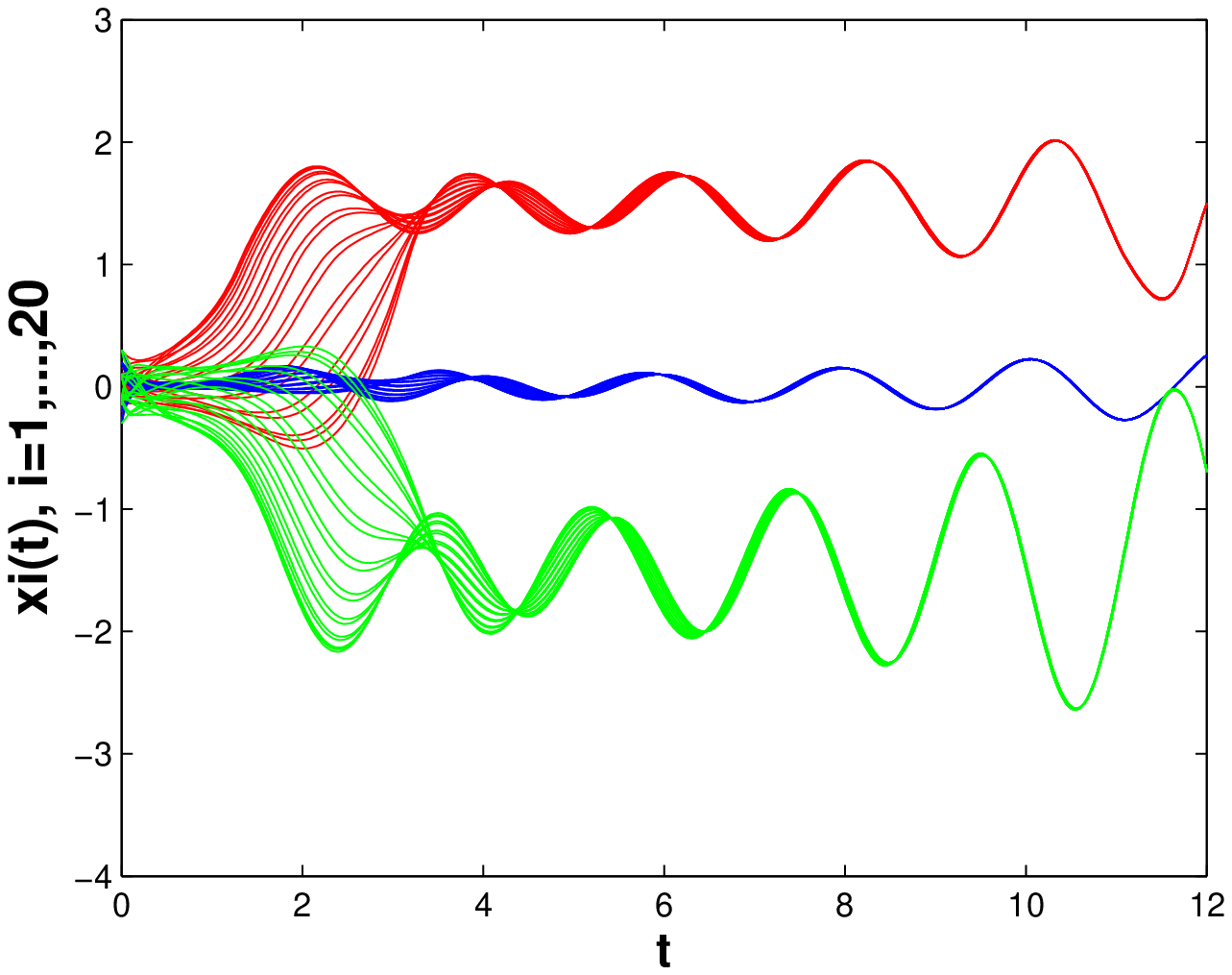} \;\quad \epsfxsize5cm
\epsfysize5cm \epsffile{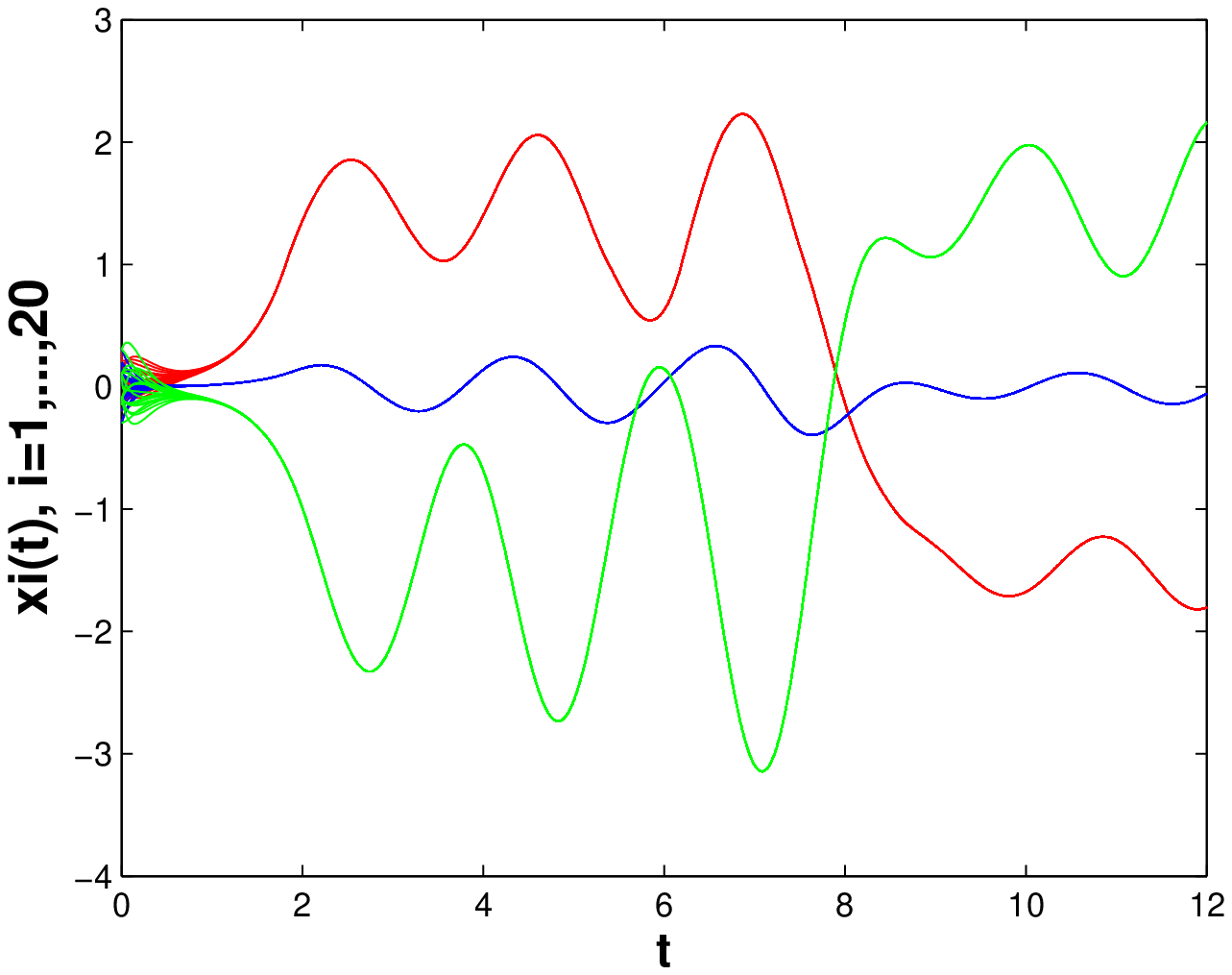}\;\quad
\epsfxsize5cm \epsfysize5cm \epsffile{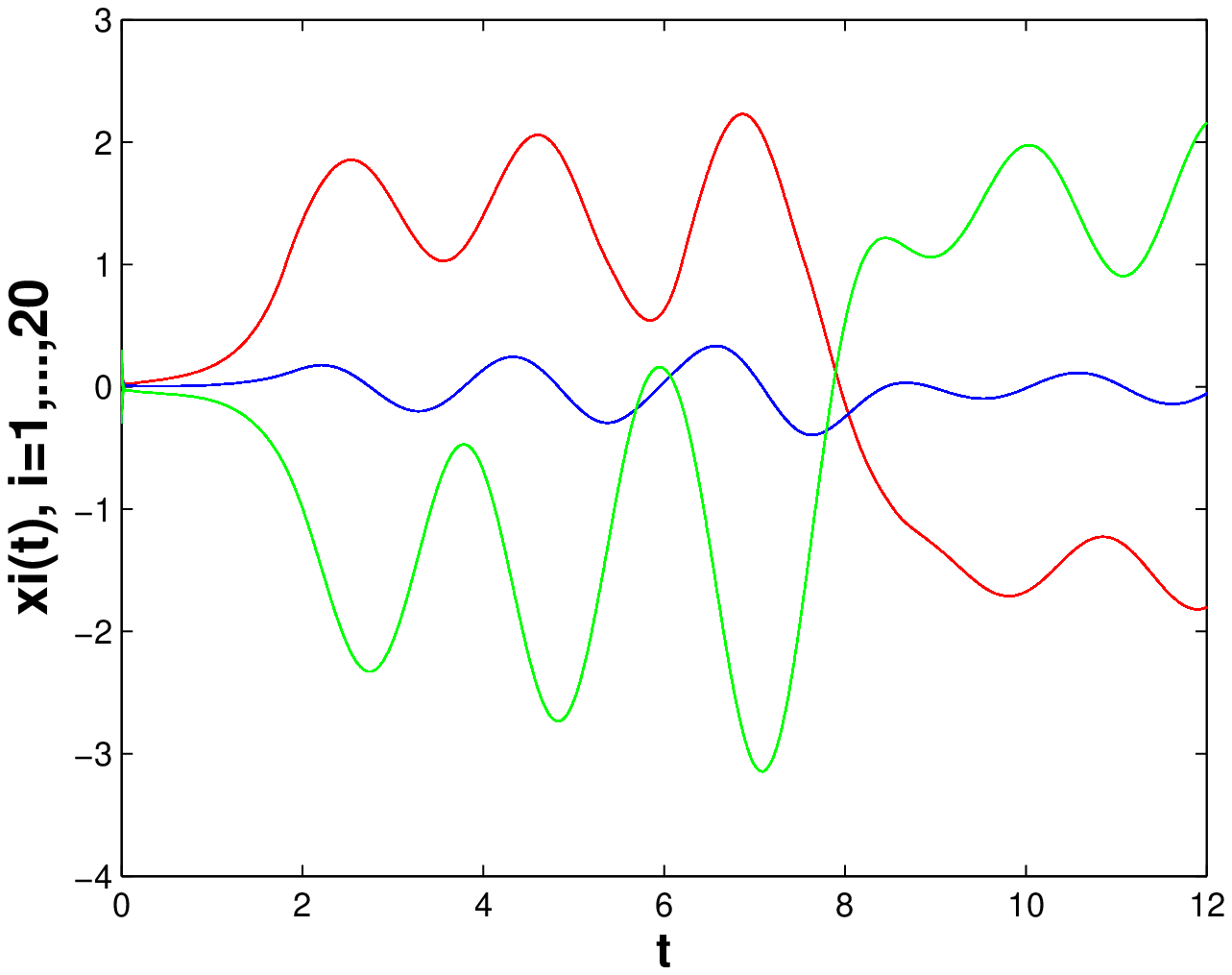}}
\end{center}

\vskip -0.7cm\quad\; {\small Nearest-neighbor coupling
}\qquad\qquad\quad {\small Star-shaped coupling
}\qquad\qquad\qquad\; {\small Global coupling}

\begin{center}
\vskip -0.5cm
 \unitlength=1cm
 \qquad \hbox{\hspace*{0.1cm} \epsfxsize5cm \epsfysize5cm
\epsffile{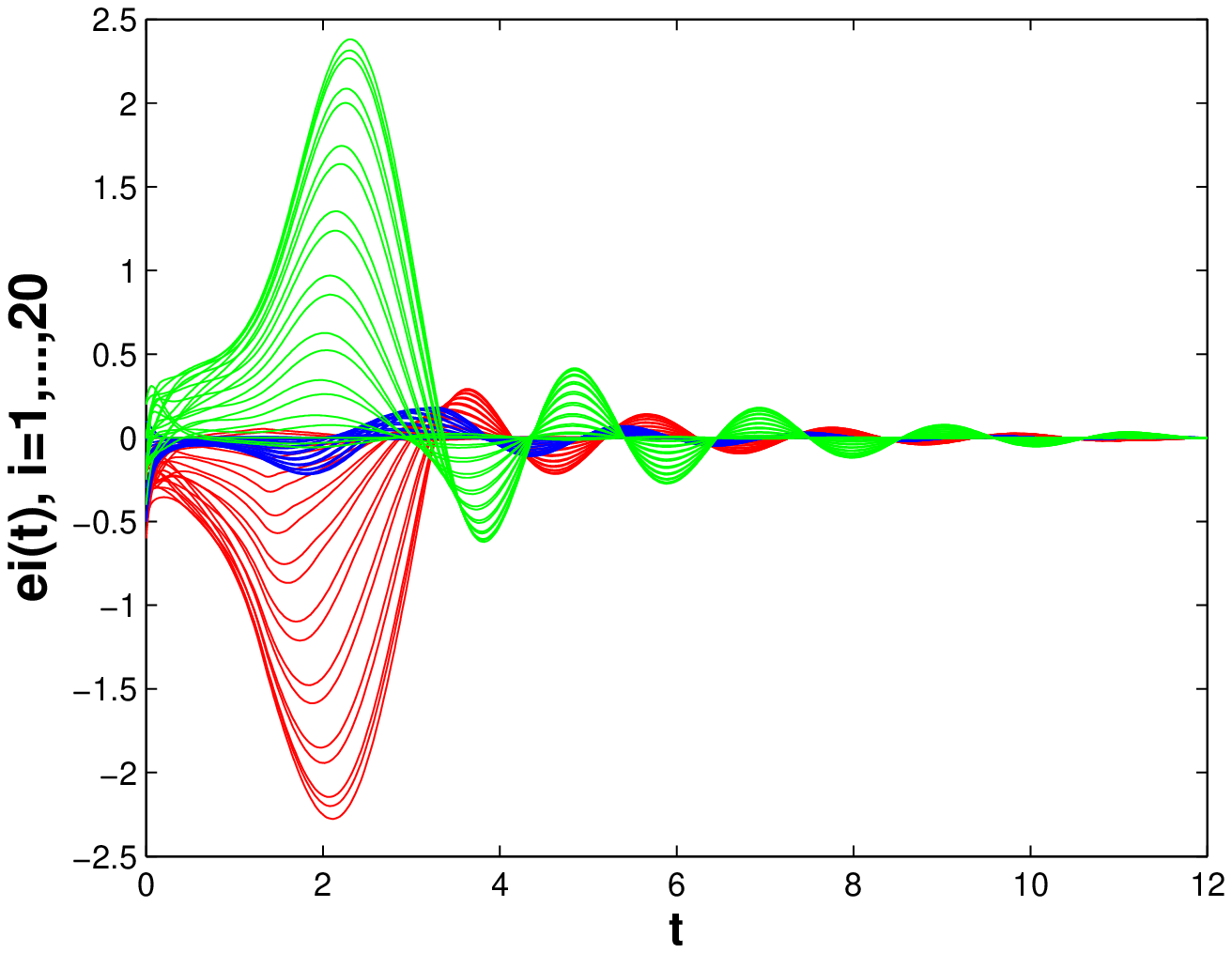} \;\quad \epsfxsize5cm \epsfysize5cm
\epsffile{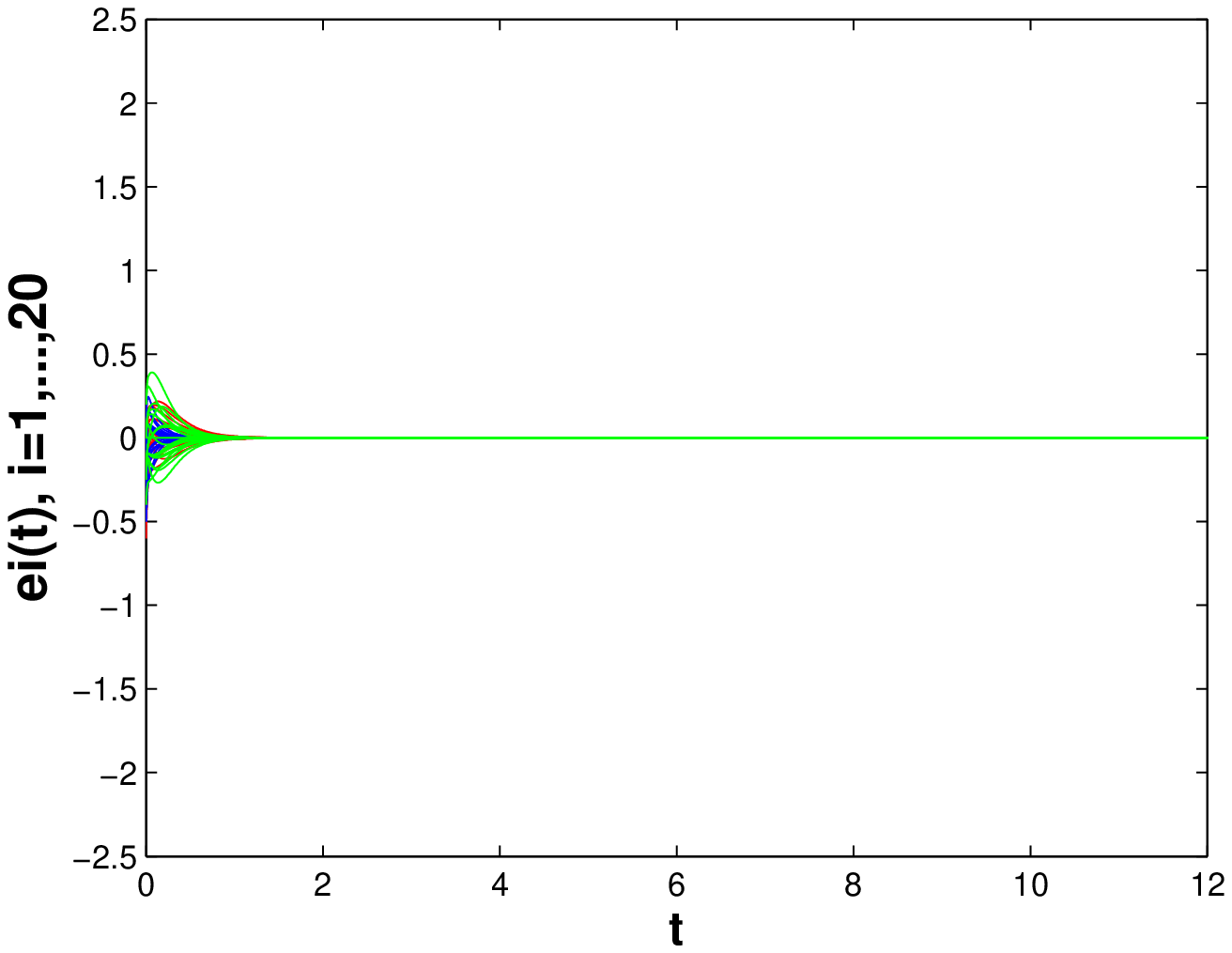}\;\quad \epsfxsize5cm \epsfysize5cm
\epsffile{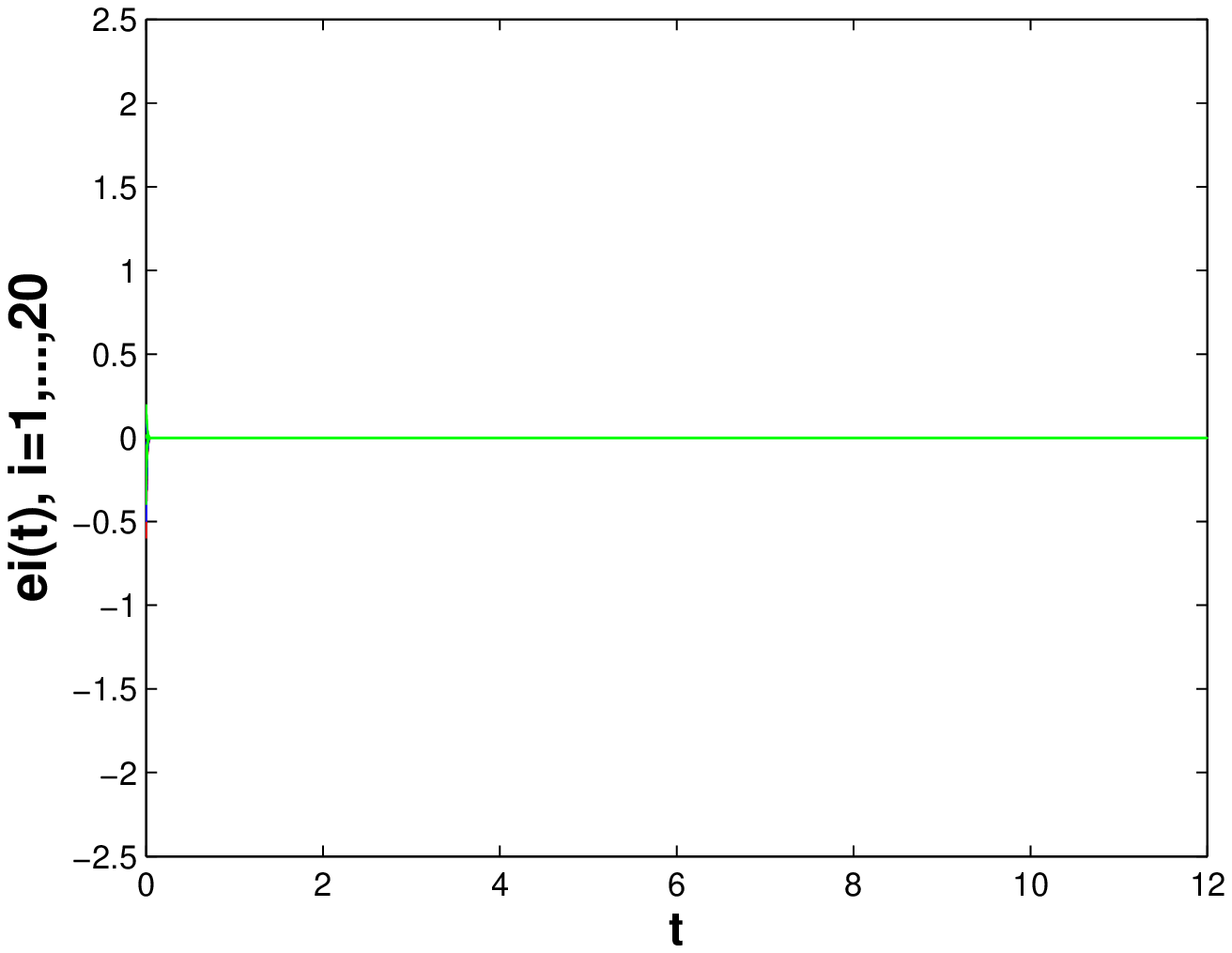}}
\end{center}

\vskip -0.7cm\quad\; {\small Nearest-neighbor coupling
}\qquad\qquad\quad {\small Star-shaped coupling
}\qquad\qquad\qquad\; {\small Global coupling}

\qquad\quad {\small Fig. 1 \quad Synchronization of the state
variables of network \dref{net1} with 20 nodes and \vskip
-0.2cm\qquad\qquad\qquad\quad  the same coupling strength in
different coupling configurations.}

\begin{center}
\vskip 0.2cm
\begin{tabular}{|c|c|c|c|}\hline
Coupling & Nearest-neighbor & Star-shaped & Global
\\\hline $\|e\|_2$ & 9.2744 & 0.8030 & 0.4142
\\\hline
\end{tabular}
\vskip 0.2cm \quad {\small Table 1 \quad Values of $\|e\|_2$. }
\end{center}

As to the case of equilibrium synchronization of network
\dref{net1}, since the synchronous state $s$ is known, the error
vector can be defined in the following way:
\begin{equation}\label{chua4}
\begin{array}{l}
e_i=x_i(t)-s,\;i=1,\cdots,N,\\
e(t)=[e_1^T \;e_2^T \;\cdots \;e_N^T]^T.
\end{array}
\end{equation}

\textbf{Example 2:} Consider a Lur'e system,
\begin{equation}\label{lure1}
\left\{\begin{array}{l}\dot{x}_1=(A_1-2B_1C_1)x_1+B_1f_1(y_1),\\
y_1=C_1x_1,\end{array}\right.
\end{equation}
where $x_1$ is the state, $y_1$ is the measured output,
$$
A_1=\left[\begin{array}{cc}0 & 1\\ -a & -b
\end{array}\right],\; B_1=\left[\begin{array}{c}b_1 \\ b_2
\end{array}\right],\; C_1=\left[\begin{array}{cc}c_1 & c_2
\end{array}\right],\; \Gamma=\left[\begin{array}{cc}1 & 0 \\ 0 & 1
\end{array}\right],
$$
and the nonlinear function $f_1(y_1)=|y_1+1|-|y_1-1|$.

A network with system \dref{lure1} as individual nodes is given as
follows:
\begin{equation}\label{lure2}
\left\{\begin{array}{l}\dot{x}=(I_N\otimes (A_1-2B_1C_1)-\sigma M\otimes \Gamma)x+(I_N \otimes B_1)f(y),\\
y=(I_N \otimes C_1)x,\end{array}\right.
\end{equation}
where $x=(x_1^T,\cdots ,x_N^T)^T$, $y=(y_1,\cdots ,y_N)^T$ and
$f(y)=(f_1(y_1), \cdots ,f_N(y_N))^T$.

Network \dref{lure2} can be viewed as a large-scale system with
measured output and feedback. Assume that system \dref{lure2} is
observable. Then, the states of network nodes achieve
synchronization if and only if all the outputs of network nodes
achieve synchronization. Thus, one only needs to consider the
outputs of network \dref{lure2}.

Replace $x_i$ with $y_i$ and set $s=0$ in \dref{chua4}. Then,
$e_i=y_i$ and $e=y$. Let $a=10,\;b=3,\;b1=0,\;b2=1,\;c_1=c_2=1,$ and
$\sigma=2$ in \dref{lure1}. Fig. 2 shows the different performances
of outputs of network \dref{lure2} with the same coupling strength
$\sigma$ ($\sigma$=2) but different coupling configurations. The
corresponding values of $\|y\|_2$ are listed in Table 2.

\begin{center}
\vskip -0.5cm
 \unitlength=1cm
 \qquad \hbox{\hspace*{0.1cm} \epsfxsize5cm \epsfysize5cm
\epsffile{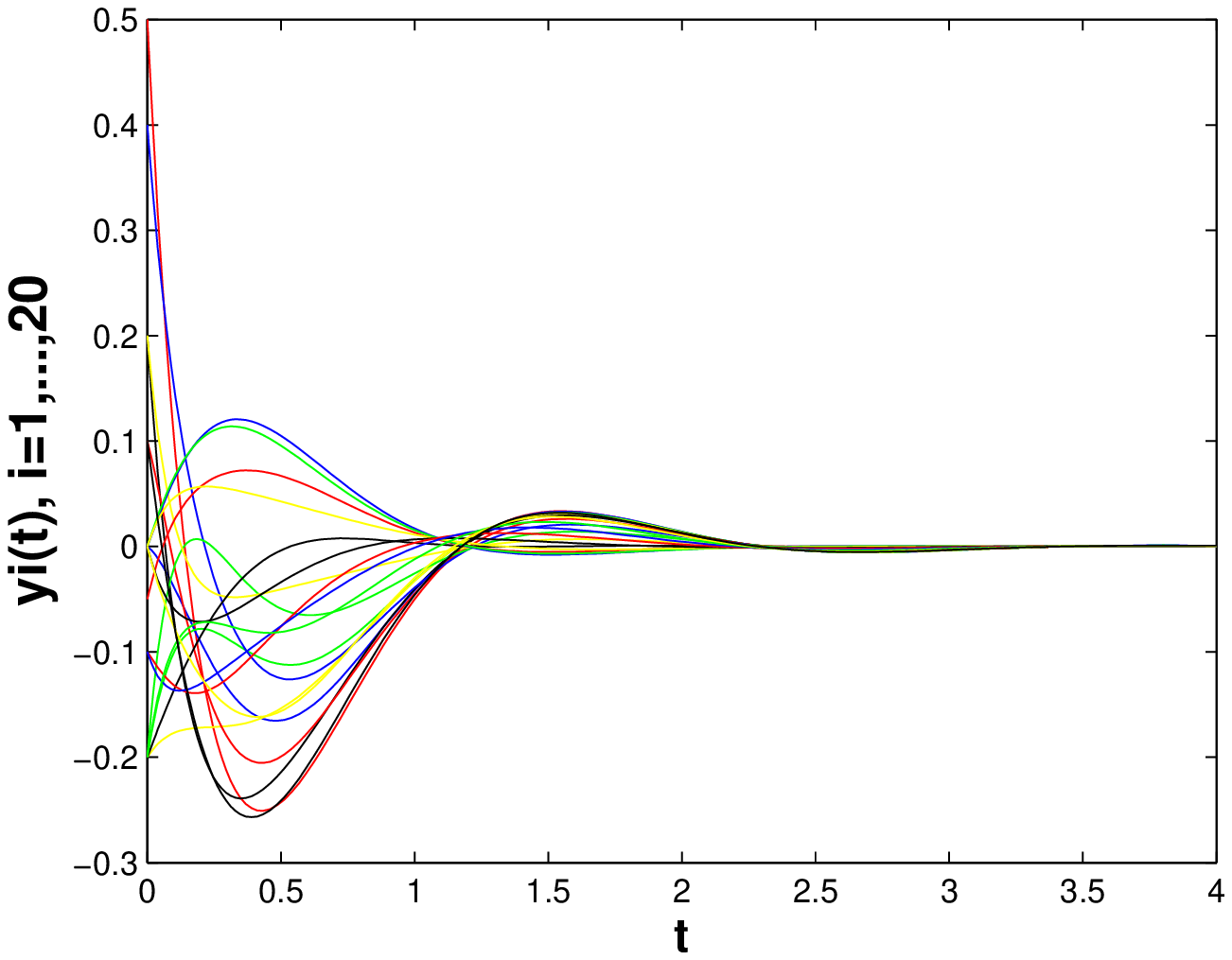} \;\quad \epsfxsize5cm
\epsfysize5cm \epsffile{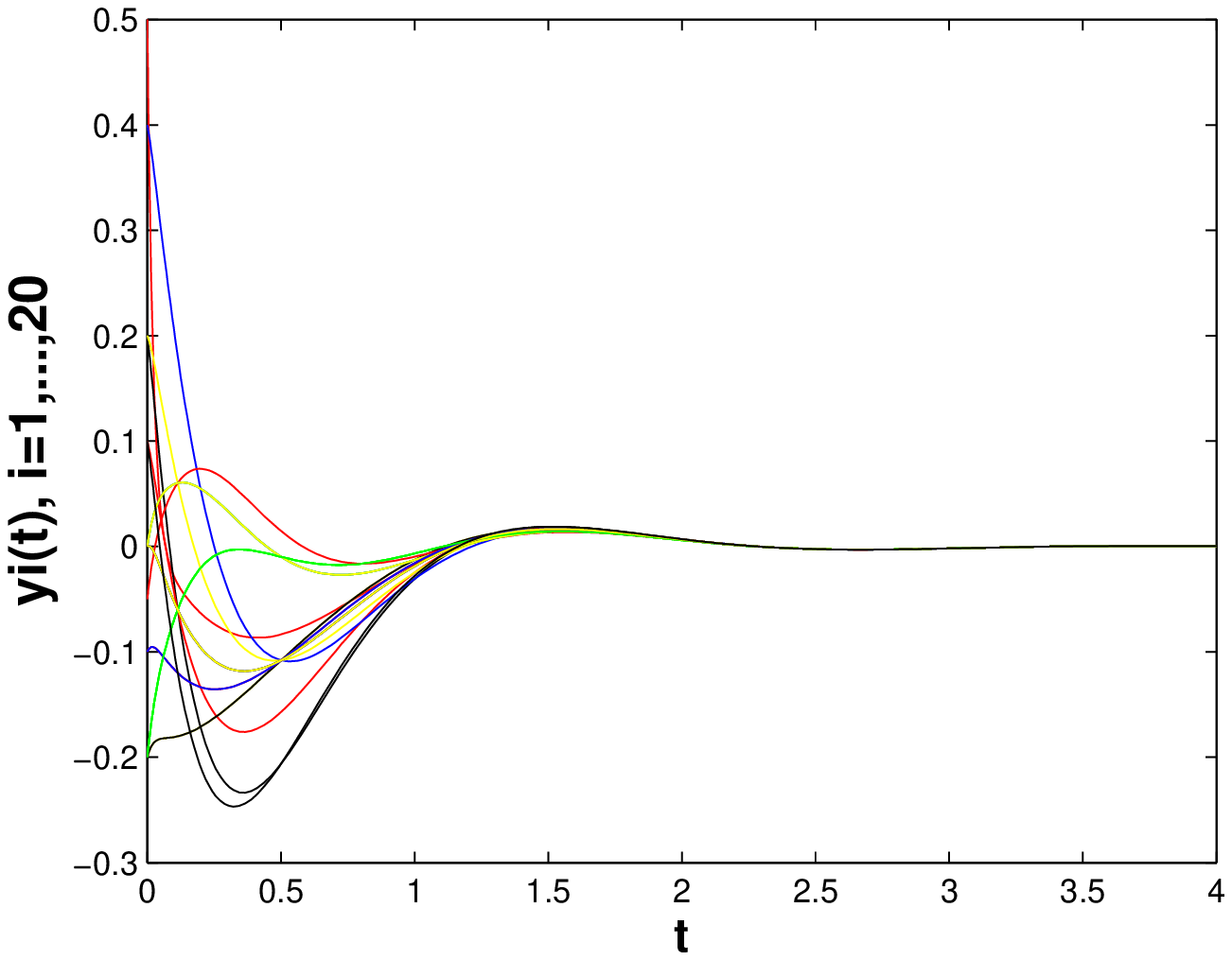}\;\quad
\epsfxsize5cm \epsfysize5cm \epsffile{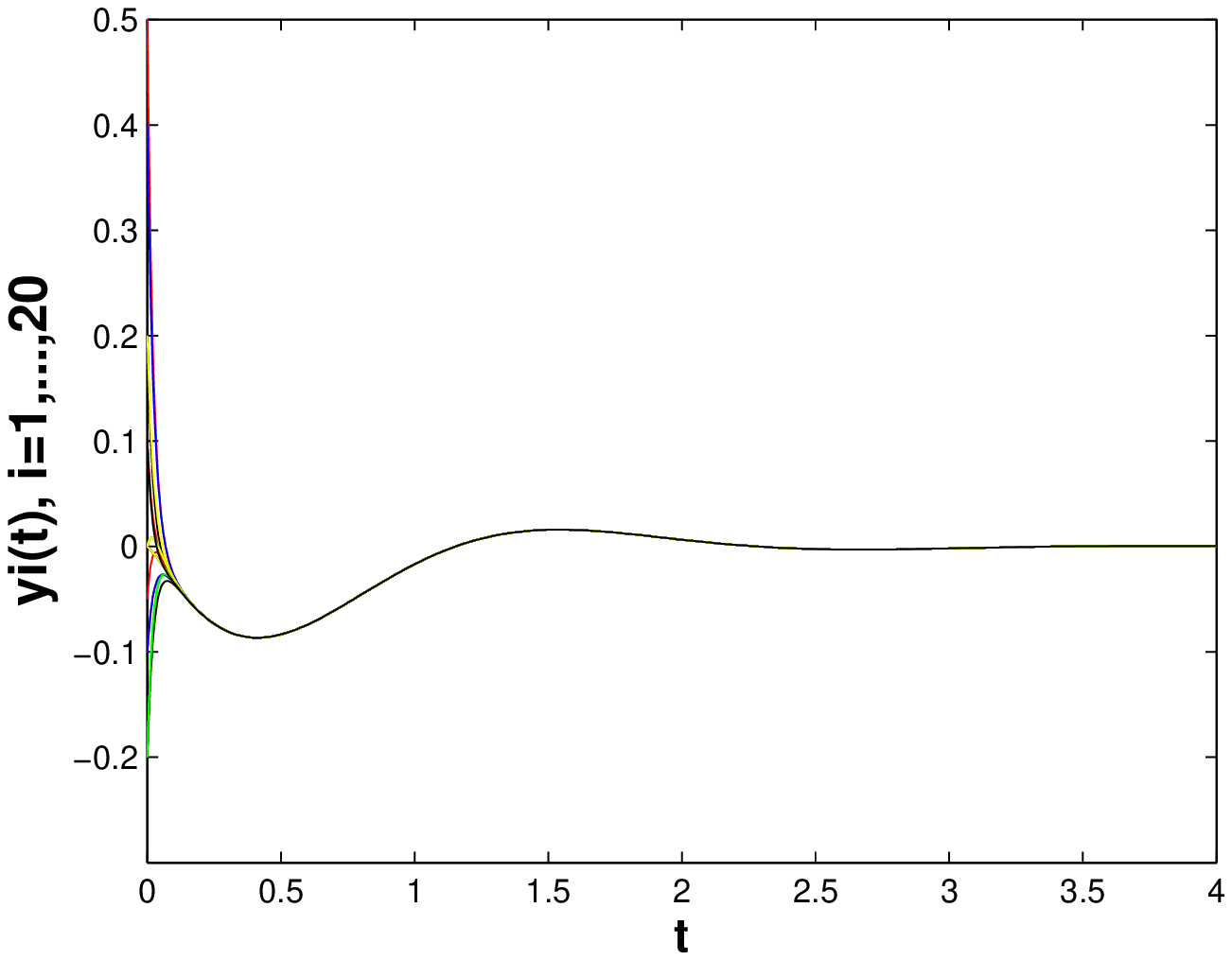}}
\end{center}

\vskip -0.7cm\quad\; {\small Nearest-neighbor coupling
}\qquad\qquad\quad {\small Star-shaped coupling
}\qquad\qquad\qquad\; {\small Global coupling}

\qquad\quad {\small Fig. 2 \quad Synchronization of the outputs of
network \dref{lure2} with 20 nodes having \vskip
-0.2cm\qquad\qquad\qquad\quad\;\;  the same coupling strength but
different coupling configurations.}

\begin{center}
\vskip 0.2cm
\begin{tabular}{|c|c|c|c|}\hline
Coupling  & Nearest-neighbor & Star-shaped & Global
\\\hline $\|y\|_2$ & 0.7296 & 0.5768 & 0.3990
\\\hline
\end{tabular}
\vskip 0.2cm \quad {\small Table 2 \quad Values of $\|y\|_2$. }
\end{center}
\hfill $\square $

\textbf{Remark 1:} As indicated by the above examples, in the same
situation of stable synchronization, the performances of
synchronization of networks can be quite different.

\textbf{Remark 2:} It is also clear that swiftness and overshoot are
important indexes for describing the synchronous behaviors. The
$L_2$ norm of the error vector $e$, i.e., $\|e\|_2$ as defined in
\dref{net3} or \dref{chua4} for the equilibrium synchronization
case, can properly measure the swiftness and overshoot of the
network synchronization: the smaller the $\|e\|_2$, the faster with
lower overshoot the network synchronization. Thus, the quantity
$\|e\|_2$ can be taken as a performance index of network
synchronizability.

\textbf{Remark 3:} The quantity $\|e\|_2$ is influenced by many
factors of the network, such as

(1) network structure, particularly the coupling strength $\sigma$
and the outer coupling matrix $M$;

(2) dynamical components, particularly the individual dynamics
determined by $f$, the synchronous state $s(t)$, and the inner
linking matrix $\Gamma$.

\section{Local synchronization to network equilibrium point}

\quad In the literature where the eigenvalues of the outer coupling
matrix are used to measure the network synchronizability, it is a
topic of great interest to investigate the relationship between the
eigenvalues and the network structure thereby finding suitable ways
to enhance the synchronizability. In this section, for the case of
equilibrium synchronization, the relationship between the quantity
$\|e\|_2$ and some network parameters is explored.
%the diagonalizability of the outer coupling matrix.

\subsection{$\|G\|_2$ as a constraint of $\|e\|_2$}

\quad Suppose that each single node in network \dref{net1} has a
measured output $y=Cx$. Then, the equations of network \dref{net1}
can be written as follows:
\begin{equation}\label{lsyn1}
\left\{
\begin{array}{l}
\dot{x}_i=f(x_i)-\sigma\sum\limits_{j=1}^Nm_{ij}\Gamma
x_j,\;x_{i}(0)=x_{i0},\\
y_i=Cx_i,
\end{array}\qquad i=1,\cdots, N.\right.
\end{equation}
If $C=I_n$, then $y_i=x_i$ is just the state of the $i$th node.

Let $s$ denote an equilibrium point of the individual node,
satisfying $f(s)=0$. Then, the linearized equations %%%%%% variational equation
 of \dref{lsyn1} about the
synchronous solutions $x_i=s$ are as follows:
\begin{equation}\label{lsyn2}
\left\{
\begin{array}{l}
\dot{\eta}_i=Df(s)\eta_i-\sigma\sum\limits_{j=1}^{N}m_{ij}\Gamma
\eta_j,\;\eta_{i}(0)=x_{i0}-s=\eta_{i0},\\
e_i=C\eta_i,
\end{array}
\qquad i=1,\cdots, N,\right.
\end{equation}
where $\eta_i$ ($\eta_i=x_i-s$) and $e_i$ are the state error vector
and the output error vector to the $i$th node, respectively. Viewing
the impulse function $\eta_{i0}\delta(t)$ as an input to system
\dref{lsyn2}, system \dref{lsyn2} can be equivalently written as
\begin{equation}\label{lsyn3}
\left\{
\begin{array}{l}
\dot{\eta}_i=Df(s)\eta_i-\sigma\sum\limits_{j=1}^{N}m_{ij}\Gamma
\eta_j+\eta_{i0}\delta(t),\;\eta_i(0_-)=0,\\
e_i=C\eta_i,
\end{array}
\qquad i=1,\cdots, N.\right.
\end{equation}
Using the Kronecker product, the error system \dref{lsyn3} can be
rewritten as
\begin{equation}\label{lsyn4}
\left\{
\begin{array}{l}\dot{\eta}=(I_N\otimes Df(s)-\sigma M\otimes \Gamma )\eta+(I_N\otimes I_n)u(t),\\
e=(I_N\otimes C)\eta,
\end{array}\right.
\end{equation}
where $\eta=[\eta_1^T \;\eta_2^T \;\cdots \;\eta_N^T]^T$, $e=[e_1^T
\;e_2^T \;\cdots \;e_N^T]^T$ and $u(t)=\eta_{0}\delta(t)$.

As in Example 2, suppose that system \dref{lsyn4} is observable.
Then, $\|e\|_2$ can be taken as a measure of the swiftness and
overshoot of the synchronization of network \dref{lsyn1}. Let $G(s)$
denote the transfer function of the error equation \dref{lsyn4} from
$u$ to $e$. Then
\begin{equation}\label{lsyn5}
G(s)=(I_N\otimes C)[s(I_N\otimes I_n)-(I_N\otimes Df(s)-\sigma
M\otimes \Gamma )]^{-1}(I_N\otimes I_n).
\end{equation}

Let the transfer function $G(s)$ be given in \dref{lsyn5}, then
\textbf{Lemma 3:} The inequality
\begin{equation}\label{lsyn6}
\|e\|_2\leq \|G\|_2\|\eta_{0}\|_2
\end{equation}
holds, where $\|e\|_2$ is the $L_2$ norm of the error vector $e(t)$
on the interval $[0,\infty)$, $\|G\|_2$ is the $H_2$ norm of the
transfer function $G(s)$, and $\|\eta_{0}\|_2$ is the vector 2-norm
of the initial error vector $\eta_{0}$.

{\bf Proof:} Since $u(t)=\eta_{0}\delta(t)$, $e(t)=g(t)\eta_{0}$,
where $g(t)$ denotes the corresponding bilateral Laplace transform
of $G(s)$. Then, by Parseval's identity,
$$
\|e(t)\|_2=\|g(t)\eta_{0}\|_2=\|G\eta_{0}\|_2\leq
\|G\|_2\|\eta_{0}\|_2.
$$
\hfill $\square $

Thus, the quantity $\|e\|_2$ is upper-bounded by the product
$\|G\|_2\|\eta_{0}\|_2$. Since $\eta_{0}$ is the given initial error
vector, $\|G\|_2$ can be taken as a constraint of $\|e\|_2$. In
fact, as introduced in Sec.1, $\|G\|_2^2$ equals the overall output
energy of the system response to the impulse input.

\textbf{Remark 4:} The advantages of using the quantity $\|G\|_2$
include:

(1) $\|G\|_2$ can be numerically computed;

(2) the synchronizability is affected by many factors of a network,
while $\|G\|_2$ can be seen as an overall reflection of these
network factors consisting of both structural and dynamical ones.

\textbf{Example 3:} Consider Example 2 again. The linearized
equation of network \dref{lure2} about the equilibrium point $s=[0\;
0\; 0]$ is given as follows:
\begin{equation}\label{lsyn7}
\left\{\begin{array}{l}\dot{\eta}=(I_N\otimes A_1-\sigma M\otimes
\Gamma)\eta
+(I_N \otimes I_n)\eta_0\delta(t),\;\;\eta(0_-)=0,\\
e=(I_N \otimes C)\eta.\end{array}\right.
\end{equation}

The corresponding values of $\|G\|_2$ with the three different
network configurations are listed in Table 3.

\begin{center}
\vskip 0.2cm
\begin{tabular}{|c|c|c|c|}\hline
Coupling  & Nearest-neighbor & Star-shaped & Global
\\\hline
$\|G\|_2$ & 2.5091 & 2.4383 & 1.2565
\\\hline
$\|e\|_2$ & 0.7296 & 0.5768 & 0.3990
\\\hline
\end{tabular}
\vskip 0.2cm \quad {\small Table 3 \quad  Values of $\|G\|_2$ of
network \dref{lure2} with different network configurations.}
\end{center}

\subsection{Relationship between $\|G\|_2$ and network structure}

\quad Example 3 shows that $\|G\|_2$ is influenced by the network
configurations. Thus, the relationship between $\|G\|_2$ and the
network structure is a problem deserving further investigation.

In this section, it is always assumed that the outer coupling matrix
$M$ is symmetrical. Then, there exists a unitary matrix $U\in
\mathbb{R}^{N\times N}$ such that $M=U\Delta U^{-1}$, where
$\Delta=\textrm{diag}(\lambda_1,\lambda_2,\cdots,\lambda_N)$ is a
diagonal matrix with the diagonal entries
$\lambda_i,\;i=1,\cdots,N$, being the eigenvalues of matrix $M$.
%$0=\lambda_1<\lambda_2\leq\cdots\leq\lambda_N$.
Let $\eta=(U\otimes I_n)\xi$, $u=(U\otimes I_n)\omega$,
 and $z=(U^{-1}\otimes I_l)e$.
Then, the error equation \dref{lsyn4} can be rewritten as follows:
\begin{equation}\label{analyH1}
\left\{
\begin{array}{l}\dot{\xi}=(I_N\otimes Df(s)-\sigma \Delta\otimes \Gamma )\xi+(I_N\otimes I_n)\omega,\\
z=(I_N\otimes C)\xi.
\end{array}\right.
\end{equation}
Note that system \dref{analyH1} is composed of $N$ uncoupled
subsystems:
\begin{equation}\label{analyH2}
\left\{
\begin{array}{l}
\dot{\xi}_i=(Df(s)-\sigma \lambda_i\Gamma)\xi_i+I_n\omega_i,\\
z_i=C\xi_i,
\end{array}\qquad i=1,\cdots, N,\right.
\end{equation}
where $\omega_i$ is the $i$th component of the input vector
$\omega=(U^{-1}\otimes I_n)u=(U^{-1}\otimes I_n)\eta_0\delta(t)$.
Let
$$
E_i=\textrm{diag}(0,\cdots,0,1,0,\cdots,0),
$$
be a matrix with the $i$th diagonal entry being 1 and all the other
entries being zero. Then, $\omega_i=(E_i\otimes I_n)(U^{-1}\otimes
I_n)\omega=(E_iU^{-1}\otimes I_n)\eta_0\delta(t)$ and
$\omega_{i0}=(E_iU^{-1}\otimes I_n)\eta_0$.

From \dref{analyH2}, the condition for ensuring the stable
synchronous state
$$
x_1=x_2=\cdots =x_N=s
$$
is that the $N$ matrices
\begin{equation}\label{lsyn7}
Df(s)-\sigma\lambda_i\Gamma,\qquad i=1,\cdots, N,
\end{equation}
are all stable.

Let $T(s)$ denote the transfer function of \dref{analyH1} from
$\omega$ to $z$. Then
$$
\begin{array}{rl}
 T(s) & =(I_N\otimes C)[s(I_N\otimes I_n)-(I_N\otimes Df(s)-\sigma
\Delta\otimes \Gamma )]^{-1}(I_N\otimes I_n)\\
& =(U^{-1}\otimes I_{l})G(s)(U\otimes I_{m}).
\end{array}
$$
Since both $(U^{-1}\otimes I_{l})$ and $(U\otimes I_{m})$ are
unitary transformations, one has
\begin{equation}\label{analyH3}
\|T(s)\|_{2}=\|G(s)\|_{2}.
\end{equation}
Let $T_i(s)=C[sI_n-(Df(s)-\sigma \lambda_i \Gamma )]^{-1}I_n$ denote
the transfer function of the $i$th subsystem in \dref{analyH2}. Then
$$
T(s)=\left[\begin{array}{ccc}T_1(s) & &
\\ &\ddots &  \\ & & T_N(s)\end{array}\right]
$$
and
\begin{equation}\label{analyH4}
\|T(s)\|_{2}^2=\sum\limits_{i=1}^{N}\|T_i(s)\|_{2}^2.
\end{equation}

\textbf{Assumption 1:} Suppose that the outer coupling matrix $M$ is
symmetrical, diffusive and irreducible, with the off-diagonal
entries $m_{ij}\leq 0$ and the diagonal entries
$m_{ii}=-\sum_{j=1,j\neq i}^Nm_{ij}$, for $i,\;j=1,2,\cdots,N$.

The outer coupling matrix $M$ satisfying Assumption 1 has a zero
eigenvalue of multiplicity $1$, and its other eigenvalues are all
positive.

\textbf{Theorem 1:} Suppose that the inner linking matrix
$\Gamma=kI$, where $k>0$ is a constant and $I$ the identity matrix.
Suppose that the outer coupling matrix $M$ satisfies Assumption 1
with the eigenvalues given as follows:
\begin{equation}\label{analyH5}
0=\lambda_1<\lambda_2\leq\cdots\leq\lambda_N.
\end{equation}
Then
\begin{equation}\label{analyH6}
\|T_{i}(s)\|_{2}^2\leq\|T_{j}(s)\|_{2}^2,
\end{equation}
where $1\leq j<i\leq N$, and
\begin{equation}\label{analyH7}
\begin{array}{l}
\|T(s)\|_{2}^2\geq \|T_1(s)\|_{2}^2+(N-1)\|T_N(s)\|_{2}^2\geq
N\|T_N(s)\|_{2}^2,\\
\|T(s)\|_{2}^2 \leq\|T_1(s)\|_{2}^2+(N-1)\|T_2(s)\|_{2}^2\leq
N\|T_1(s)\|_{2}^2.
\end{array}
\end{equation}

{\bf Proof:} Let $P$ be an arbitrary positive definite matrix such
that
$$
P(Df(s)-\sigma \lambda_j \Gamma)+(Df(s)-\sigma \lambda_j
\Gamma)^TP+C^TC<0.
$$
By the assumption that $\Gamma=kI$, one has
$$
P\Gamma+\Gamma^TP>0.
$$
Then
$$
\begin{array}{rl}
& P(Df(s)-\sigma \lambda_{i} \Gamma)+(Df(s)-\sigma \lambda_i
\Gamma)^TP+C^TC\\
& =P(Df(s)-\sigma \lambda_j \Gamma)+(Df(s)-\sigma \lambda_j
\Gamma)^TP+C^TC-\sigma(\lambda_i-\lambda_{j})(P\Gamma+\Gamma^TP)\\
& <-\sigma(\lambda_i-\lambda_{j})(P\Gamma+\Gamma^TP)\leq0.
\end{array}
$$
The above inequality and Lemma 1 together leads to the assertions of
the theorem. \hfill $\square $

\textbf{Remark 5:} Since $\lambda_1=0$, $\|T_1(s)\|_{2}$ in
\dref{analyH7} represents the $H_2$ norm of the transfer function of
each individual node in the network.

\textbf{Remark 6:} Theorem 1 provides a simple way for comparing
$\|G\|_{2}$ of a network with its different possible structures. It
can be deduced from Theorem 1 that $\|G\|_2$ will not increase as
the eigenvalues of the outer coupling matrix increase.

\textbf{Example 4:} Consider Example 2 again. The eigenvalues of the
global coupling matrix $M_{glo}$ are $\lambda_1=0$ and
$\lambda_2=\cdots=\lambda_N=N$, while the eigenvalues of the
star-shaped coupling matrix $M_{sta}$ are $\lambda_1=0$,
$\lambda_2=\cdots=\lambda_{N-1}=1$ and $\lambda_N=N$. Let $G_{glo}$
and $G_{sta}$ denote the transfer functions of the global coupling
and the star-shaped coupling networks, respectively. Then, by
Theorem 1,
$$
\begin{array}{rl}
\|G_{glo}\|_{2}^2= \|T_{glo}\|_{2}^2
& =\|T_1(s)\|_{2}^2+(N-1)\|T_{\lambda=N}\|_{2}^2\\
& \leq
\|T_1(s)\|_{2}^2+\|T_{\lambda=N}\|_{2}^2+(N-2)\|T_{\lambda=1}\|_{2}^2\\
& = \|T_{sta}\|_{2}^2= \|G_{sta}\|_{2}^2.
\end{array}
$$
It is consistent with the numerical results given in Example 3.

For Example 2, through integral computations, an analytical
expression of $\|T_i\|_{2}^2$ can be obtained as follows:
\begin{equation}\label{analyH8}
\|T_i\|_{2}^2=\frac{2(c_1^2+c_2^2)\sigma^2\lambda_i^2+[(3c_1^2+c_2^2)b+2c_1c_2(1-a)]\sigma\lambda_i
+(c_1b-c_2a)^2+(c_1^2+c_2^2)a+c_1^2}
{4\sigma^3\lambda_i^3+6b\sigma^2\lambda_i^2+(4a+2b^2)\sigma\lambda_i+2ab}.
\end{equation}
By \dref{analyH8} and \dref{analyH4}, Table 4 gives the different
values of $\|G_{glo}\|_{2}$ and $\|G_{sta}\|_{2}$ of network
\dref{lure2} as the node number $N$ increases.

\begin{center}
\begin{tabular}{|c|c|c|c|c|c|c|}\hline
N & 1 & 10 & 100 & 1000 & $\infty$ \\\hline $\|G_{glo}\|_{2}$ &
1.0801 & 1.3190 & 1.4492 & 1.4696  & 1.2910
\\\hline $\|G_{sta}\|_{2}$ & 1.0801 & 2.2619 & 6.9840 & 22.0434 & $\infty$ \\\hline
\end{tabular}
\vskip 0.2cm \quad {\small Table 4 \quad Values of $\|G\|_2$ of
network \dref{lure2} with different network configurations.}
\end{center}
\hfill $\square$

\section{Optimal controller design}

\quad So far, the pinning control strategy is extensively used for
 achieving synchronization of dynamical networks
\cite{li04,wang02}. The main advantage of the pinning control
strategy is that only a few network nodes are needed to be
controlled. AS far as the control effects (referring to the
swiftness and overshoot of the synchronization) and the control cost
(represented by the $L_2$ norm of the control input) are concerned,
however, pinning control may not be the best choice. It is revealed
by the \textit{LQR} problem, as illuminated in \dref{pre6}, when
matrices $C$ and $D$ are properly selected, both the control effects
and the control cost can be simultaneously evaluated by the $L_2$
norm of the measured output $y$ of the controlled network. The
smaller the $\|y\|$, the better the control effects and the lower
the control cost. In this section, based on the optimal solution to
the \textit{LQR} problem, an \textit{LQR} optimal controller is
designed for network \dref{net1}, which can drive the network onto
some homogenous stationary states while minimizing the quantity
$\|y\|$.
% in terms of the linearized network systems.

Suppose that the controlled network is given as follows:
\begin{equation}\label{opticon1}
\left\{\begin{array}{l}
\dot{x}_i=f(x_i)-\sigma\sum\limits_{j=1}^{N}m_{ij}\Gamma
x_j+Bu_i\\
y_i=Cx_i+Du_i,
\end{array}\qquad i=1,\cdots, N,\right.
\end{equation}
where $B\in \mathbb{R}^{n\times m}$, $C\in \mathbb{R}^{l\times n}$,
$D\in \mathbb{R}^{l\times m}$ are given constant matrices,
$u=(u_1^T,\cdots,u_N^T)^T$ is the controller to be designed, and
$y=(y_1^T,\cdots,y_N^T)^T$ is the measured output of the LQR
problem. As in Sec. 4, let $s$ denote an equilibrium point of the
individual node and let $\eta=(\eta_1^T,\cdots,\eta_N^T)^T,$ where
$\eta_i=x_i-s,\;i=1,\cdots, N,$ be the state error vector.

Suppose that the controller is a feedback law of the state error,
i.e., $u=F\eta$, where $F$ is the feedback gain matrix to be
determined. Then, by the Kroneck product, the equation of the
linearized system of \dref{opticon1} about the synchronous solution
$x_1=\cdots=x_N=s$ is given as follows:
\begin{equation}\label{opticon2}
\left\{\begin{array}{l}\dot{\eta}=A_c\eta+B_cu,\\
y=C_c\eta+D_cu,
\end{array}\right.
\end{equation}
where $A_c=I_N\otimes Df(s)-\sigma M\otimes \Gamma $,
$B_c=I_N\otimes B$, $C_c=I_N\otimes C$ and $D_c=I_N\otimes D$.
Furthermore, suppose that the matrices $A_c,\;B_c,\;C_c$ and $D_c$
satisfy the assumptions (A1-A4) as given in Lemma 2. Then, by Lemma
2, the feedback gain $F$ for the \textit{LQR} problem of the
linearized system \dref{opticon2} is obtained as
\begin{equation}\label{opticon3}
F=-(B_c^*X+D_cC_c),
\end{equation}
where $X$ is the stabilizing solution to the following Riccati
equation:
\begin{equation}\label{opticon4}
(A_c-B_cD_c^*C_c)^*X+X(A_c-B_cD_c^*C_c)-XB_cB_c^*X+C_c^*(D_c)_\bot
(D_c)_\bot^*C_c=0.
\end{equation}

\textbf{Remark 7:} The optimal controller $u=F\eta$ with $F$ given
in \dref{opticon3} has the property that it controls network
\dref{opticon1} to the synchronous state $x_1=\cdots=x_N=s$ and
meanwhile minimizes the $L_2$ norm of the output $y$ of the
linearized system \dref{opticon2}.

As the node number $N$ of network \dref{opticon1} increases, the
direct computation of the feedback gain $F$ as given in
\dref{opticon3} will become harder. For the case when the outer
matrix $M$ is symmetrical, the feedback gain $F$ can be
alternatively given in terms of the $N$ uncoupled subsystems. The
main advantage of this approach is that the decentralized feedback
gain $F_i$ will be given only based on the information of the $i$th
subsystem, $i=1,2,\cdots,N$.

By using the unitary transformation, as used in Sec. 4, system
\dref{opticon2} becomes
\begin{equation}\label{opticon5}
\left\{
\begin{array}{l}\dot{\xi}=(I_N\otimes Df(s)-\sigma \Delta\otimes \Gamma )\xi+(I_N\otimes B)\omega,\\
z=(I_N\otimes C)\xi+(I_N\otimes D)\omega,
\end{array}\right.
\end{equation}
where $\eta=(U\otimes I_n)\xi$, $u=(U\otimes I_n)\omega$,
 and $z=(U^{-1}\otimes I_l)y$. The $N$ uncoupled controlled subsystems
 are as follows:
\begin{equation}\label{opticon6}
\left\{
\begin{array}{l}
\dot{\xi}_i=A_{i}\xi_i+B\omega_i,\\
z_i=C\xi_i+D\omega_i,
\end{array}
\qquad i=1,\cdots, N,\right.
\end{equation}
where $A_{i}=Df(s)-\sigma \lambda_i\Gamma$. By the analysis in Sec.
4, one has
$$
\|y\|_2^2 =\|z\|_2^2 =\sum_{i=1}^{N}\|z_i\|_2^2.
$$

Suppose that the constant matrices $A_{i},\;B,\;C$ and $D$ in
\dref{opticon6} satisfy the assumptions (A1-A4) as given in Lemma 2,
for $i=1,\cdots,N.$ Then, the controller $F_i$ to the $i$th
subsystem can be given as follows:
\begin{equation}\label{opticon7}
F_i:=-(B^*X_i+DC),
\end{equation}
where $X_i$ is the stabilizing solution to the following Riccati
equation:
\begin{equation}\label{opticon8}
(Df(s)-\lambda_i \Gamma-BD^*C)^*X_i+X_i(A-\lambda_i
\Gamma-BD^*C)-X_iBB^*X_i+C^*D_\bot D_\bot^*C=0,
\end{equation}
for $i=1,2,\cdots,N$. Moreover, the feedback gain matrix $F$ is
given by
\begin{equation}\label{opticon9}
F=(U\otimes I_m)\left[\begin{array}{ccc}F_1 & &
\\ &\ddots &  \\ & & F_N\end{array}\right](U^*\otimes
I_n).
\end{equation}

\textbf{Example 5:} Consider Example 1 again. Suppose that the
network is with the global coupling configuration. The objective is
to design a controller such that network \dref{net1} with Chua's
oscillators can be driven to the synchronous solution
$x_1=\cdots=x_N=s=[0\;0\;0]^T$.

It is easy to deduce that
$$Df(s)=\left[\begin{array}{ccc}-\alpha(m_2+1) & \alpha & 0\\ 1 & -1 & 1\\
0 & -\beta & -\gamma\end{array}\right].$$ Consequently, the
linearized equation of the controlled network is given by
\begin{equation}\label{exam52}
\left\{\begin{array}{l}\dot{\eta}=(I_N\otimes Df(s)-\sigma M\otimes \Gamma )\eta+(I_N\otimes B)u,\\
y=(I_N\otimes C)\eta+(I_N\otimes D)u,
\end{array}\right.
\end{equation}
where the constant matrices $B=C=D=I_3$ in this example.

By the above analysis, the \textit{LQR} optimal controller can be
designed by solving the Riccati equation \dref{opticon4}. For
comparison, a pinning control strategy is also considered, where $6$
nodes are randomly selected to be pinned by controllers
$u_i=-\sigma\cdot d\cdot I_3\eta_i$ with $\sigma=1$ and $d=20$.

Fig. 3 shows the different performances of the synchronous behavior
of the two controlled networks, by the \textit{LQR} optimal
controller and the pinning controller, respectively. The
corresponding values of $\|y\|_2$ are listed in Table 5.

\begin{center}
%\vskip -0.5cm
 \unitlength=1cm
 \qquad \hbox{\hspace*{0.1cm} \epsfxsize7cm \epsfysize5cm
\epsffile{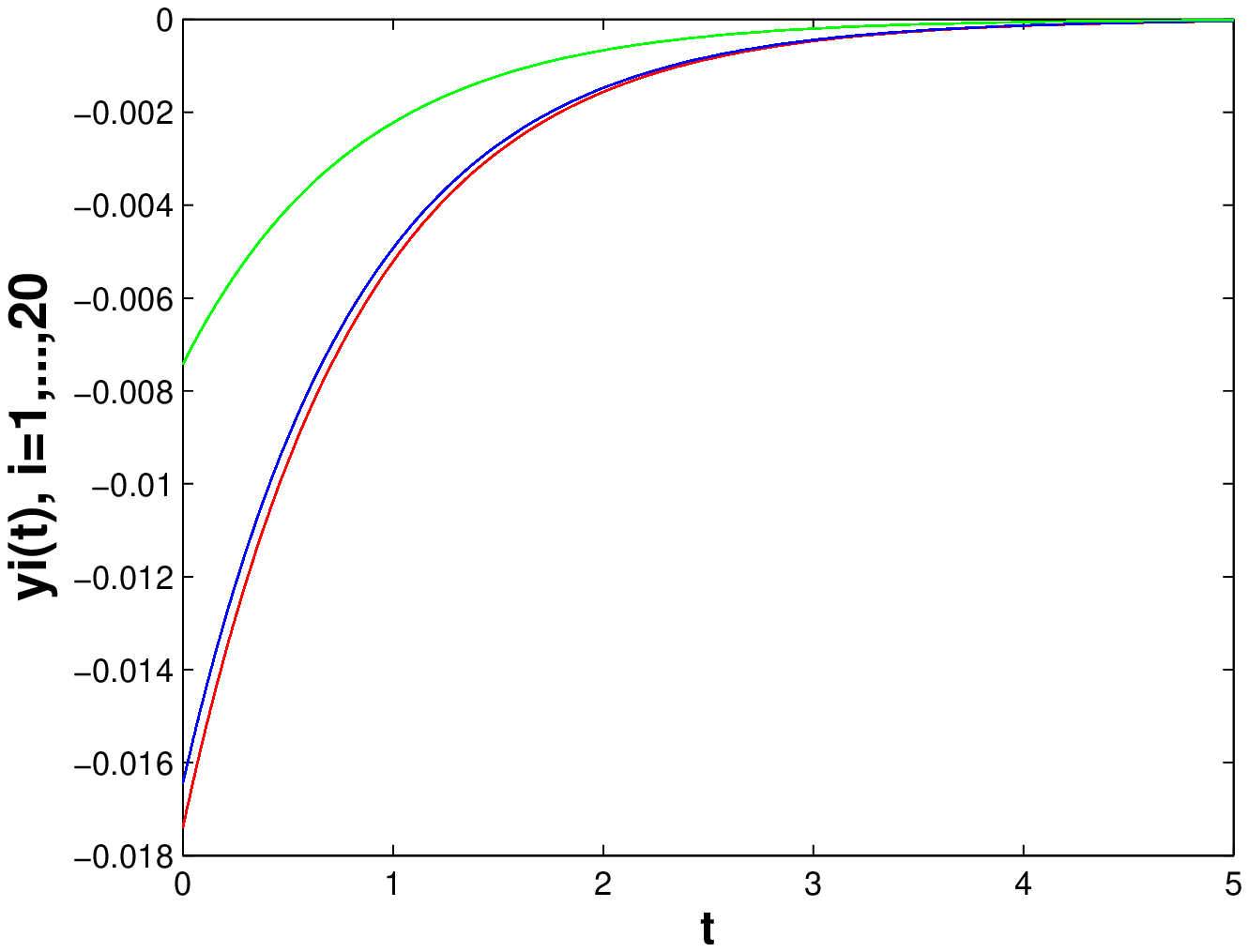} \;\quad \epsfxsize7cm \epsfysize5cm
\epsffile{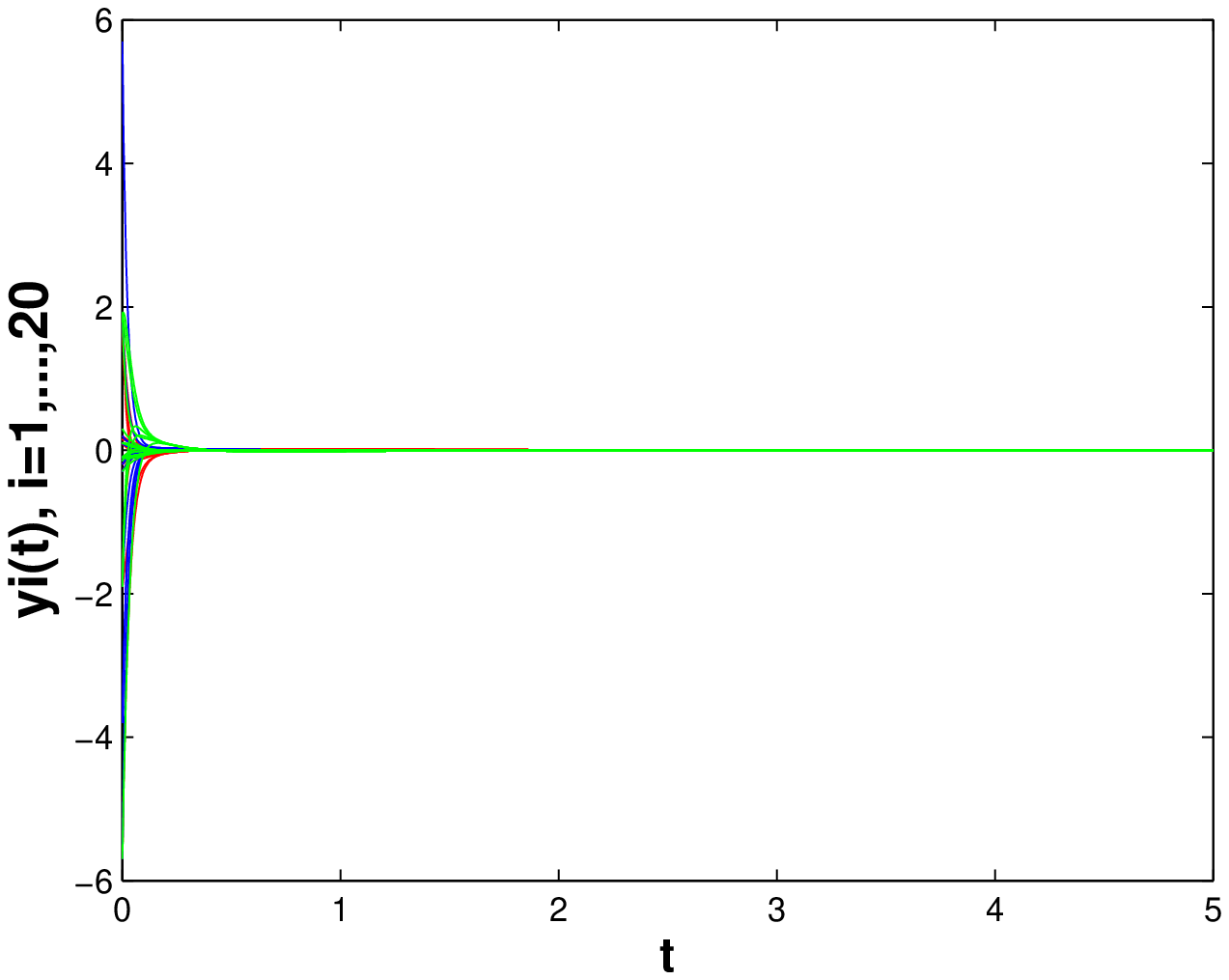}}
\end{center}

\vskip -0.7cm\qquad\qquad\qquad\qquad\; {\small \textit{LQR} optimal
control}\qquad\qquad\qquad\qquad\qquad\qquad\; {\small pinning
control}

\qquad\qquad\qquad\qquad\qquad {\small Fig. 3 \quad Different
effects of the \textit{LQR} and pinning controllers.}

\begin{center}
\vskip 0.2cm
\begin{tabular}{|c|c|c|c|}\hline
Controller & \textit{LQR} optimal control & Pinning control
\\\hline $\|y\|_2$ & 0.1309 & 4.9799

\\\hline
\end{tabular}
\vskip 0.2cm \quad {\small Table 5 \quad Values of $\|y\|_2$. }
\end{center}
\hfill $\square$

\section{Conclusion}

\quad In this paper, synchronizability of dynamical networks is
considered based on some new measures: the swiftness and overshoot
of the network synchronization. The quantity $\|e\|_2$, which
represents the $L_2$ norm of the synchronization error vector
$e(t)$, is taken as the performance index of this kind of
synchronizability. It has been shown by several numerical examples,
$\|e\|_2$ presents a suitable measure of both swiftness and
overshoot of network synchronization: the smaller the values of
$\|e\|_2$, the faster with smaller overshoot the network
synchronization. For the case when the synchronous state is an
equilibrium point, $\|e\|_2$ is upper-bounded by the product of the
vector 2-norm of the initial error vector $e_0$ and the $H_2$ norm
of the transfer function $G(s)$, denoted as $\|G(s)\|_2$, of the
linearized network about the equilibrium point. The relationship
between $\|G(s)\|_2$ and the network structure has also been
discussed. Under some assumptions, it has been proved that
$\|G(s)\|_2$ will not increase as the real eigenvalues of the outer
coupling matrix increase. Finally, based on the techniques of the
\textit{LQR} control theory, an optimal controller has been
suggested to drive the network onto some homogenous stationary
states, which , in the mean time, can minimize the $L_2$ norm of the
output of the linearized network. Further research along this
direction seems to be quite promising as long as the network energy
and performance are concerned, therefore deserves further efforts.

%\section*{\large Acknowledgements}
%\quad This work is supported by both the National Science
% Foundation of China under grants 60674093, 60334030 and the City University of Hong
%Kong under the Research Enhancement Scheme and SRG grant 7002134.

\end{document}